# Building LOFAR – status update

**M.A. Garrett**[1,2,3,4]

[1]*ASTRON, Netherlands Institute for Radio Astronomy, Postbus 2, 7990AA, Dwingeloo, The Netherlands*
[2]*Sterrewacht Leiden, Postbus 9513, 2300 RA Leiden, The Netherlands*
[3]*Centre for Astrophysics and Supercomputing, Swinburne University of Technology, Australia*
[4]*On behalf of the LOFAR Consortium.*

*E-mail:* `garrett@astron.nl`

The Low Frequency Array (LOFAR) is a new generation of electronic radio telescope based on aperture array technology and working in the frequency range of 30-240 MHz. The telescope is being developed by ASTRON, and currently being rolled-out across the Netherlands and other countries in Europe. The plan is to build at least 36 stations in the Netherlands (with baseline lengths of up to 100 km), 5 stations in Germany, and 1 station in each of Sweden, France and the UK. With baseline lengths of up to 2000 km, sub-arcsecond resolution will be possible at the highest frequencies. The Key Science Projects being addressed by the project include: deep, wide-field cosmological surveys, transients, the epoch of re-ionisation and cosmic ray studies. We present the current status of the project, including the development of the super-core in Exloo and the completion of the first 3 stations. "First fringes" from these stations is also presented.

## 1. Introduction to LOFAR

ASTRON, the Netherlands Institute for Radio Astronomy, is developing and building LOFAR (the Low Frequency Array) – an ambitious, next generation radio telescope that fully exploits recent advances in digital signal processing, fibre-based distributed communication networks and high performance super-computing. Operating in a largely unexplored region of the electro-magnetic spectrum (30-240 MHz), LOFAR will consist of a distributed interferometric array of dipole antenna stations that have no moving parts, and that permit large areas of the sky to be imaged simultaneously – up to 8 independent fields of view can be instantaneously steered to any position on the sky using electronic beam-forming techniques. Over the next year, at least 36 stations will be built in various locations across the Netherlands. The densely populated core comprising 13 antennas is located in Exloo, not far from ASTRON's head-quarters in Dwingeloo. The longest baseline within the Netherlands is ~ 100 km. At least another 8 stations will be located in 4 other European countries – 5 in Germany, and 1 each in Sweden, France & the UK. There is also significant interest in Italy, Poland, Ukraine and Ireland, but these stations are not yet fully funded. The physical distribution of stations (including potential stations) is shown in Figure 1.

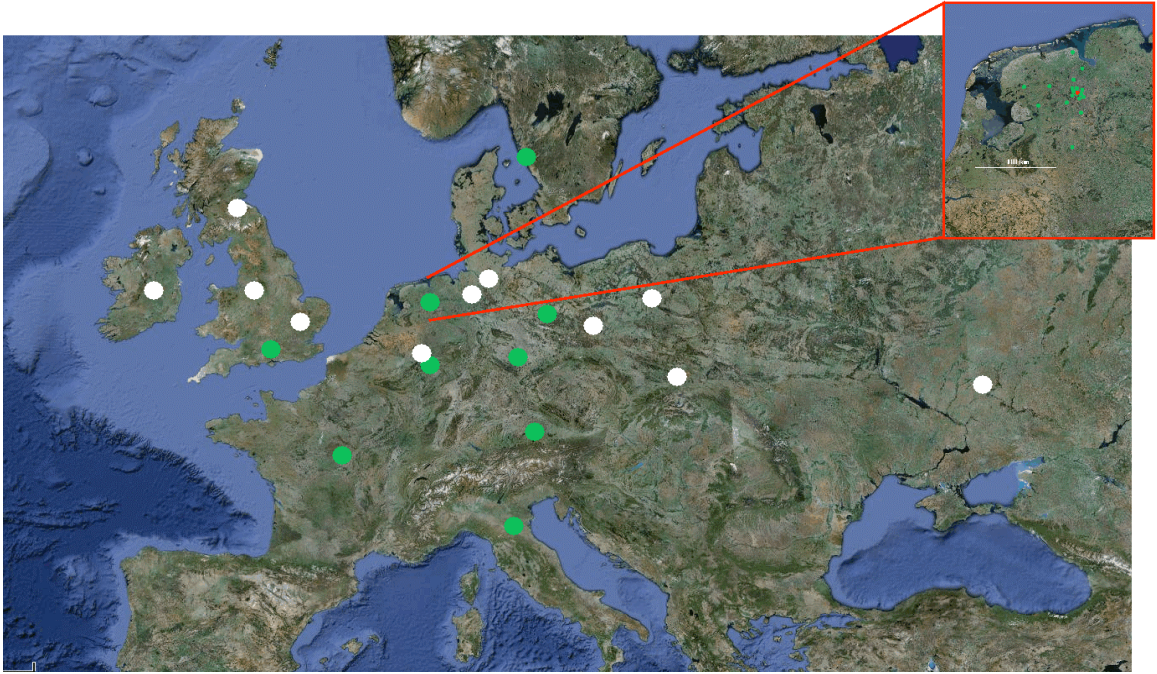

*Figure 1: The location of LOFAR stations in Europe and in the Netherlands. Green dots represent stations where funding is secured or for which letters of intent have been received. White dots represent stations that are planned by the various international consortia and for which funding is being sought. The location of remote antennas in the Netherlands is shown top right – the red dot shows the location of the central core stations in Exloo.*

LOFAR uses two types of antenna: the low-band antenna (LBA, 15-80 MHz) and the high-band antenna (HBA 110-240 MHz). The Dutch LOFAR stations are actively populated with ~ 48 dual polarisation LBA elements and 48 (2 x 24) HBA tiles (each HBA tile is composed of 16 dual-polarisation antenna elements). In fact at each station, 96 LBA elements are physically present, but only 48 are active at any one time – the exact configuration employed depends on the observing frequency. A typical station physically spans an area of about 180 x 100 m$^2$ across. To ensure good sensitivity on the longer baselines, the international stations have twice as many antenna elements as the Dutch stations (see de Vos et al. 2009 for a complete system description). An image of the first complete station, CS302 is presented in Figure 2.

LOFAR represents a step-change in the evolution of radio astronomy technology. It offers many new modes of observation, including an "all-sky" transient detection capability. By buffering large amounts of data at the individual dipole level, retrospective imaging of the entire sky is possible. LOFAR's ability to playback buffered data, together with the telescopes huge field of view, ushers in a new era for radio astronomy. In addition to the study of transient phenomena, the LOFAR Key Science Programmes (KSP) include: (i) deep, wide-field extra-galactic surveys; (ii) exploring the epoch of re-ionisation, and (iii) cosmic ray detection. Both the survey and transient KSPs are likely to discover many interesting phenomena that will require much higher-resolution observations with e-VLBI. When completed in 2010, the full LOFAR array will have a sensitivity and resolution that is ~ 2 orders of magnitude better than anything that has proceeded it, and will open up one of the last unexplored frontiers of observation astronomy.

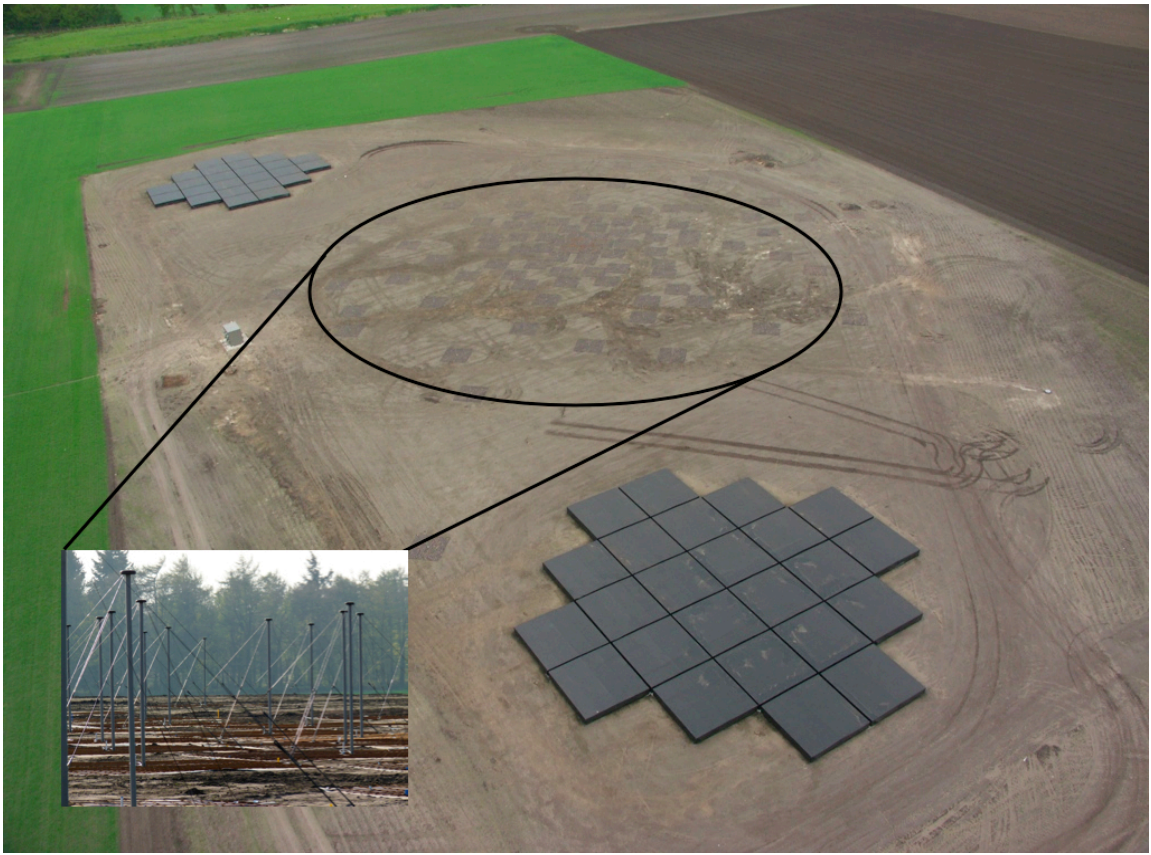

*Figure 2: The first complete station CS302. This station lies about 2 km from the super-core (see Figure 3).*

## 2. Building LOFAR

LOFAR is the first example of a "mass-production" radio telescope – the NL array alone consists of ~ 30000 HBA dipoles and 3500 LBA dipoles. Many of the lessons learned in the process of constructing LOFAR will be relevant to future radio telescope projects, especially the Square Kilometre Array (SKA).

Since late summer 2008, large areas of land in the area around Exloo and beyond have been prepared in order to raise and level the fields where LOFAR stations were to be located (see Figure 3). Unfortunately, a combination of heavy rainfall and hard frost during the winter months, made it impossible to realise the flatness required in all of the fields (~ 3 cm r.m.s.). It was therefore impossible to deploy any of the stations until a period of dry weather began in March/April. In the meantime, in order to minimise long-term delays, storage facilities were organised for the LBA and in particular the HBA antennas as they continued to roll-off the factory production lines. In the same vein, it was decided not to populate and test the electronic cabinets in the field (as originally planned) but to do this in the production line. This requires the fully laden cabinets to be transported from the factories, and deployed in the fields. So far this procedure has worked well with the sealed RFI properties of the cabinets being unaffected by the deployment process.

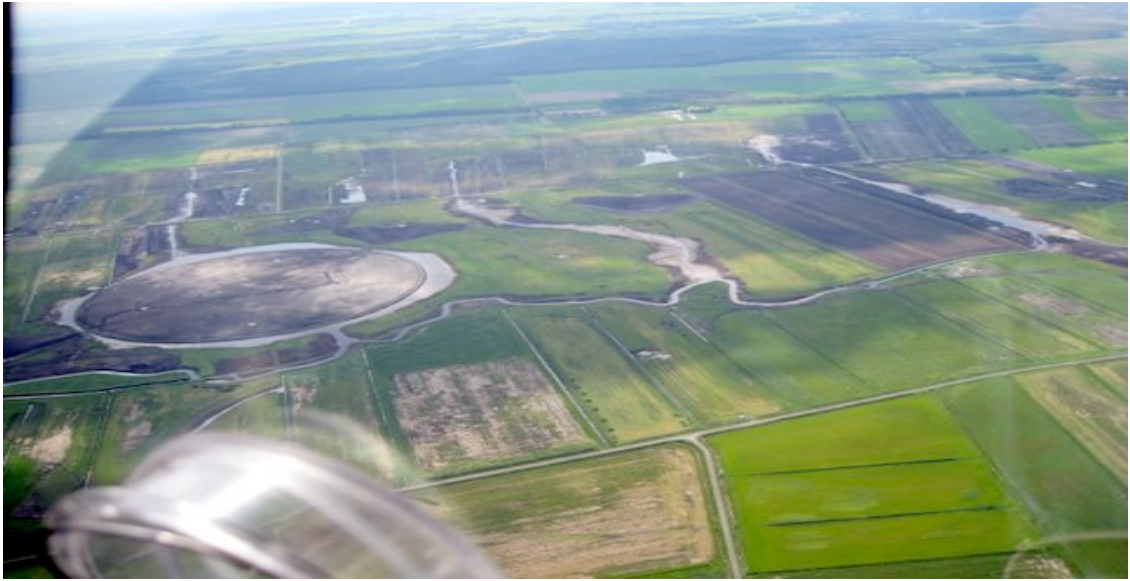

*Figure 3: Aerial photograph of the LOFAR core area at Exloo, NL. The "superterp" is shown to the left - it spans ~ 350 metres across and is raised about 1.6 metres above the local ground level. Six densely packed LOFAR stations are currently under construction there, and should be complete in October 2009. ASTRON owns 400 ha of land in the core region, and in collaboration with the provincial government, and local land & water authorities, this will be developed into a nature preservation area.*

In the spring of 2009, rapid progress was made and in the beginning of May the first station (CS302) was completed. However, around this time a new problem began to emerge, specifically the breeding season for birds threatened to halt all activities in and around Exloo. Fortunately, excellent communications with the regional government, the local population and the relevant bird protection organisations averted this nightmare scenario, and strict safeguards were introduced to ensure the well-being of nesting birds and their chicks while the station roll-out continued. Figure 4 shows some of the steps involved in realising LOFAR.

## 3. Commissioning LOFAR

With the first stations complete, tests to detect fringes began in mid-June 2009. After several weeks of testing with no clear results, first fringes were eventually discovered on the 8/9[th] of July 2009. By observing and tracking Cygnus-A for over 10 hours, the LOFAR astronomical commissioning team (led by Michiel Brentjens, ASTRON) detected first fringes to the LOFAR core station CS302 (& prototype station CS010) and two remote stations, RS307 & RS503. First fringes were first observed at a central frequency of 54 MHz and a bandwidth of 48 MHz (30 to 78 MHz) using the LBA systems. Fringes were later also detected using the HBAs (see Figure 5). Strong fringes were detected on all baselines, including the 20 km baseline formed by the remote stations. The data quality is outstanding. At the time of writing (July 2009), the first interferometric images using 4 stations were in the process of being made.

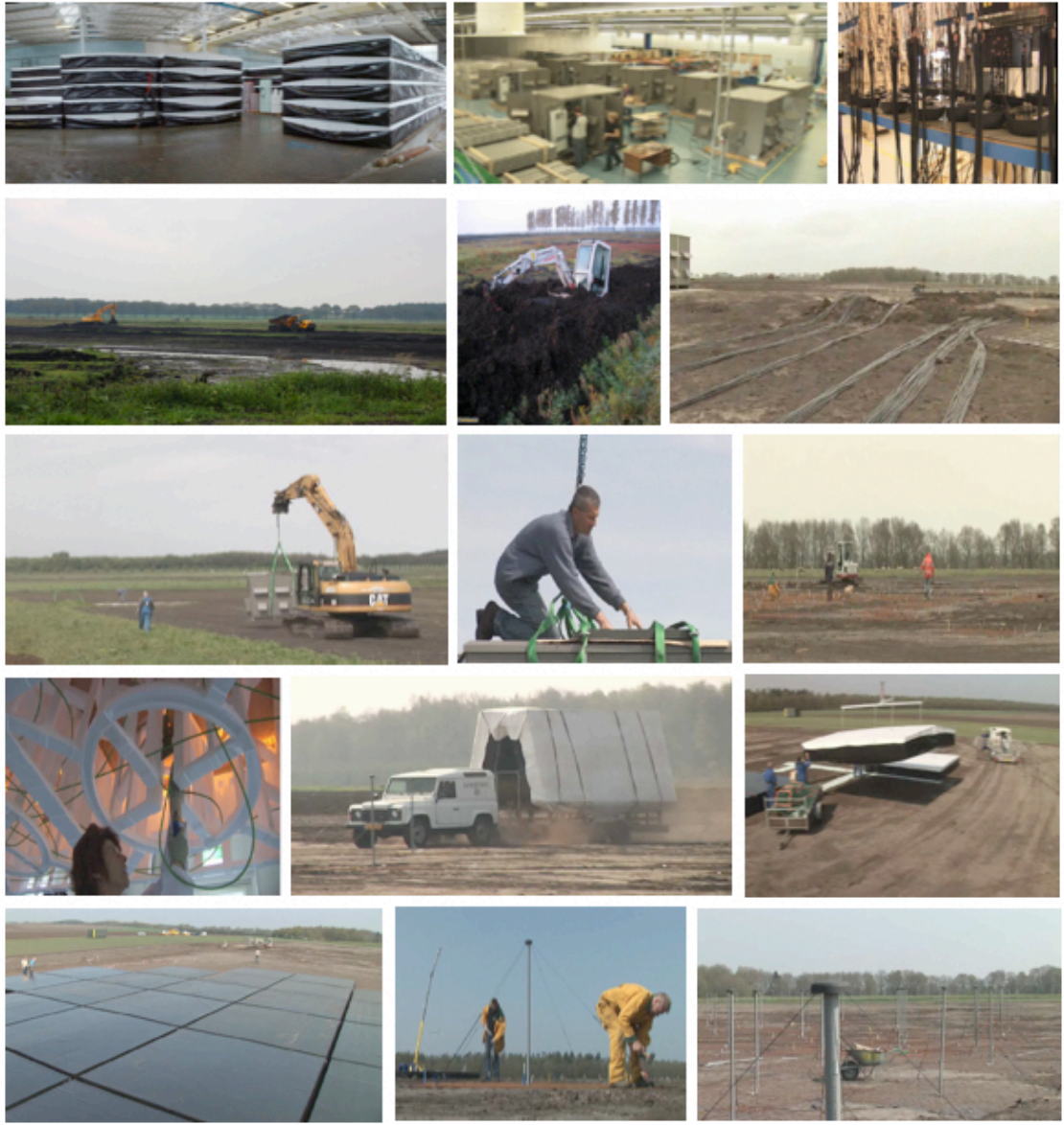

*Figure 4: Some of the stages involved in building LOFAR – producing and storing the antenna hardware, preparing the fields, placing the cabinets, digging in the cables and finally deploying the HBA and LBA antennas.*

## 4. The LOFAR International Telescope

LOFAR has evolved from a Dutch national concern, into an international radio telescope project. With the pace of progress developing ever rapidly, an Announcement of Opportunity (AO) was recently released. The AO invites both the broad international astronomical community and the KSP teams to participate in the commissioning of the telescope and to provide information on their future observing plans. The latter is required in order to scope out the future requirements of the community in terms of resources, operating model and modes of user access. A fraction of the time is expected to be dedicated to an "open skies" time allocation policy, partially funded via the RadioNet FP7 Trans-national Access programme. The governance of the International LOFAR Telescope is another issue currently under discussion. The aim is to set up an International LOFAR Consortium in which all parties are represented, including the contributing European partners. ASTRON's overall responsibility is to maximise the scientific output of the telescope and to ensure the community is well served in the broadest possible sense.

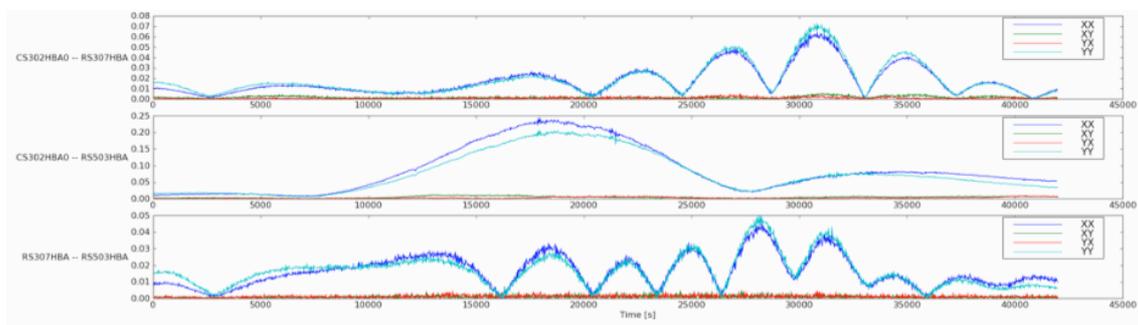

*Figure 5: LOFAR's first fringes - a plot of visibility amplitude vs time for all 4 polarisation products and the baselines involving the first 3 stations to be completed as part of the LOFAR roll-out (courtesy Michiel Brentjens, ASTRON).*

## 5. e-VLBI & LOFAR

The advances made in the field of e-VLBI over the last few years have been impressive (see these proceedings). Lessons learned regarding the connection of radio telescopes across the European continent may also be relevant to the development of the distributed International LOFAR Telescope (e.g. Garrett et al. 2009). There can be no doubt that e-VLBI and LOFAR will also be complimentary in the area of transient research. In addition, LOFAR surveys will detect a vast number of extra-galactic radio sources. The vast majority of transients and faint AGN systems will very likely remain unresolved by LOFAR, even on European baselines. Follow-up observations with higher resolution instruments, in particular e-MERLIN and e-VLBI, will be required in order to probe the morphology of these systems and to fully appreciate and understand the nature of these distant systems.